\documentclass[preprint]{aastex}

\newcommand\ione{\left< I_1 \right>}
\newcommand\itwo{\left< I_2 \right>}

\shorttitle{Asymmetric Beam Combination}
\shortauthors{Monnier}

\begin{document}

\title{Asymmetric Beam Combination for Optical Interferometry}


\author{J. D. Monnier}
\affil{Smithsonian Astrophysical Observatory MS\#42, 60 Garden Street, 
Cambridge, MA, 02138}
\email{jmonnier@cfa.harvard.edu}


\begin{abstract}
Optical interferometers increasingly use single-mode fibers as spatial
filters to convert varying wavefront distortion into intensity
fluctuations which can be monitored for accurate calibration of fringe
amplitudes.  Here I propose using an {\em asymmetric}
coupler to allow the photometric intensities of each telescope beam to
be measured at the same time as the fringe visibility, but without the
need for dedicated photometric outputs, which reduce the light throughput
in the interferometric channels.
In the read-noise limited case often encountered in the infrared, I show that a
53\% improvement in signal-to-noise ratio for the visibility amplitude
measurement is achievable, when compared to a balanced coupler setup
with 50\% photometric taps (e.g., the FLUOR experiment).  
In the Poisson-noise limit
appropriate for visible light, the improvement is reduced to only
$\sim$8\%.
This scheme also reduces the cost and complexity of the
beam combination since fewer components and detectors are required, and
can be extended to more than two telescopes for ``all-in-one" or pair-wise 
beam combination.  Asymmetric beam combination can also be 
employed for monitoring
scintillation and throughput variations in systems without spatial 
filtering.

\end{abstract}



\section{Introduction}
Spatial filtering is key for precise calibration of atmospheric
turbulence for ground-based optical interferometers.  After sidereal
delay compensation, the stellar light from each telescope is focused
onto the tip of a single-mode fiber.  Up to $\sim$78\% of the energy
\citep{shaklan88} of a perfect telescope beam can be coupled into a
single-mode fiber.  In practice, this coupling efficiency is significantly
less, depending on how well the
incoming electric field distribution overlaps with the fiber beam
profile.  This is the act of ``spatial filtering,'' whereby only the
``clean'' part of each wavefront (one spatial mode) is transmitted by the
single-mode fiber.  In essence, phase irregularities of the incoming
wavefronts are converted into amplitude fluctuations, trading a
difficult to measure quantity for one easily monitored
\citep[e.g.,][]{shaklan89}.

The light from each fiber can then be efficiently and coherently
combined in a balanced 2x2 coupler, a fiber optic component which
mixes 50\% of the light from each telescope coherently, equivalent to 
a beamsplitter.  In some cases, approximately one-half
the light is split-off from each fiber before the coupler to act as a
``photometric tap,'' a direct and real-time monitor of the coupling
efficiency for each telescope.  In this scheme the fringe amplitude
measured at the outputs of the coupler can be directly calibrated
based on the strength of the signals in the photometric taps
\citep[e.g.,][]{foresto97}.

There are two distinct ways that spatial filtering improves
calibration.  The most important effect is that the {\em mean
coupling} into each fiber is a direct measure of atmospheric
turbulence, with poor seeing resulting in low coupling efficiency.
When seeing inevitably changes in the time between observing the
target and calibrator, the coupling efficiency into the fibers will
change, and this will be seen as a change in average signal level.
However, since the interference efficiency in the 2x2 coupler is not a
function of atmospheric conditions, the overall calibration of the
system visibility is maintained.  Heterodyne detection is another kind
of spatial filtering which utilizes this principle, and is one reason
why the Infrared Spatial Interferometer \citep[e.g.,][]{hale2000} can
maintain accurate calibration despite the dearth of bright
mid-infrared calibrators.

The second way spatial filtering can improve measurement precision is
by correcting for short-term fluctuations in coupling efficiency.  The
fringe amplitude is proportional to the harmonic mean of the two
telescope intensity levels (e.g., see Eq.\,\ref{eq1}), while the DC
level is proportional to the sum of the two telescope intensity
levels.  This means that the fringe contrast, or visibility, decreases
when the signal from one telescope is significantly greater than the
other.  While this effect on the system visibility can be
statistically calibrated by using a calibrator star, the use of
photometric taps allows each individual fringe measurement to be
independently calibrated.  This second use of spatial filtering allows
high quality fringe visibility estimates with very few measurements,
since one does not have to statistically average over these
fluctuations \citep[for specific application and discussion,
see][]{foresto97}.  Note that this second aspect of spatial filtering
is of limited utility for faint sources, since it can only be applied
when the signal levels can be measured to reasonable accuracy within a
single atmospheric coherence time.

Initial experiments using spatial filters in infrared interferometry
have led to dramatic improvement in the precision of visibility
measurements.  Using flouride single-mode fibers with in-line 2x2
couplers and photometric taps, \citet{perrin99} have reported
visibility measurements with 0.3\% precision of bright M-giants in the
K-band, 10 to 20$\times$ better than the precision reported 
using traditional techniques \citep[e.g.,][]{dyck96,nordgren99}.

In this short paper, I discuss the use of {\em asymmetric} 2x2
couplers which can permit the calibration improvements derived from
photometric taps but without diverting valuable photons from the
fringe amplitude measurement.  I will detail the signal-to-noise ratio
advantages in both read-noise and Poisson-noise limiting
cases.  Lastly, I will briefly comment on how these ideas can be applied to
other beam combination schemes, such as when spatial filtering occurs
{\em after} beam combination, for pair-wise and ``all-in-one''
combiners for more than two beams, for free-space combiners with no spatial
filtering, and for fringe-tracking systems.

\section{Balanced and Asymmetric Couplers}
\begin{figure}[tbh]
\begin{center}
\includegraphics[width=4.5in]{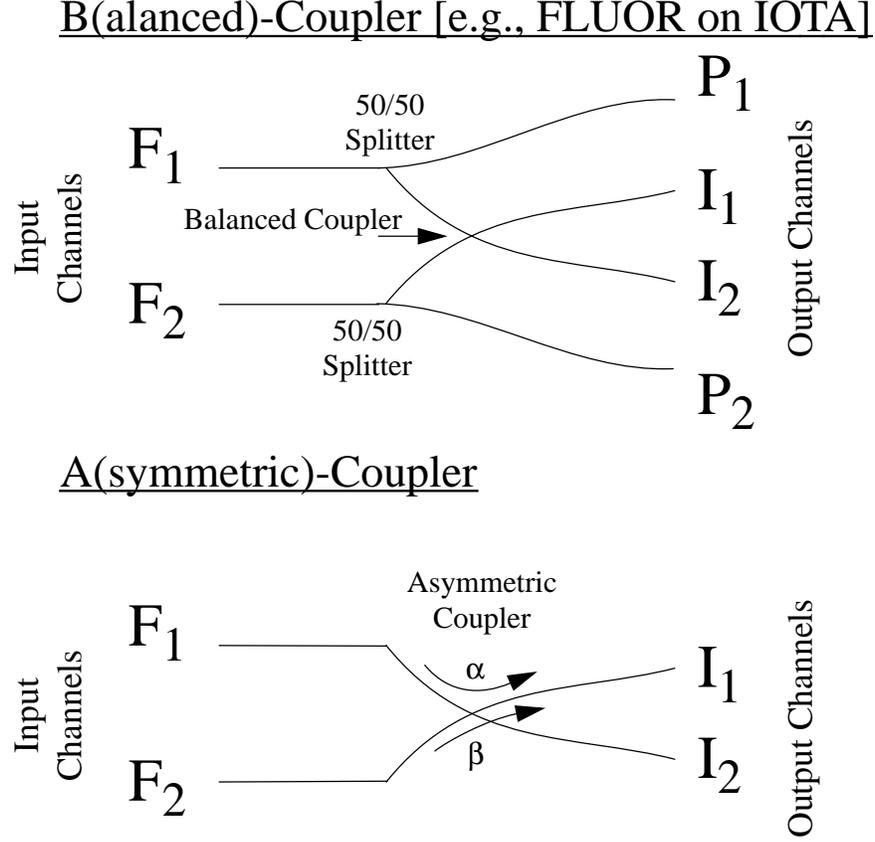}
\caption{\scriptsize This diagram illustrates the basic designs of the
FLUOR-type (Balanced) Coupler and the Asymmetric coupler design, being
proposed here.  
$F_1$ and $F_2$ represent the
incoming fluxes from telescopes~1 and 2.  {\em (top panel)}
B(alanced)-coupler design shows a 50\% photometric tap before the
remaining light from each telescope is combined in a 50/50 (balanced,
or symmetric) coupler.  $P_1$ and $P_2$ correspond to the photometric
channels, while the interferometric channels, $I_1$ and $I_2$, contain
the optical fringes.  {\em (bottom panel)} There are no photometric
channels in this design, but rather an Asymmetric coupler is used to
combine the light.  As shown in the figure, the fraction $\alpha$
($\beta$) of $F_1$ ($F_2$) appear at $I_1$, while the fraction $\beta$
($\alpha$) of $F_1$ ($F_2$) appear at $I_2$.
\label{fig:diagram}}
\end{center}
\end{figure}

As mentioned in the Introduction, current fiber optic correlators,
e.g.  FLUOR instrument on the IOTA interferometer \citep{foresto97},
VINCI on VLTI (Foresto, private communication), use a photometric tap
which removes $\sim$50\% of the light from each telescope before beam
combination in a ``Balanced'' (symmetric 50/50) coupler (hereafter,
``B-coupler'').  This is schematically indicated in the top panel of
Figure\,\ref{fig:diagram}.  However, this method is not efficient for
resolved or faint sources since roughly half the photons are being
diverted from the fringe measurement for the photometry; it is clearly
futile to monitor photometric fluctuations when the fringes can not
even be detected.

The bottom panel of Figure\,\ref{fig:diagram} shows a novel method for
recovering the photometric signals without requiring dedicated
photometric taps, thus allowing improved sensitivity while maintaining
high calibration precision.  At the center of the design is an
``Asymmetric'' coupler (hereafter, ``A-coupler'').  An A-coupler is
much like a beamsplitter manufactured to reflect a different fraction
than it transmits.  It will be shown that the photometric signals from
each telescope can be reconstructed through appropriate linear
combinations of the interferometric signals after the A-coupler,
assuming the properties of the coupler are well-known.  While the
fringe contrast from an A-coupler is slightly decreased by having
unequal contributions from each telescopes, the signal strength is
increased since no light is wasted on dedicated photometric taps.

\label{outputsigs}
For the following discussion and calculations I will refer to
Figure\,\ref{fig:diagram} which labels the various input and outputs
to the beam combiners under consideration.  The incident photon
fluxes (photons$\cdot$s$^{-1}$) coming from telescope 1 and 2 will be
denoted as $F_1$ and $F_2$ respectively.  In order to take into
account the finite quantum efficiency $\eta$ of the detector,
substitute $\eta F_i$ for $F_i$ in the subsequent formulas.  Both
combiners under consideration (``B-Coupler'' and ``A-Coupler'') have
so-called ``interferometric outputs,'' labeled $I_1$ and $I_2$, while
the ``B-Coupler'' also has two photometric outputs $P_1$ and $P_2$.

For this calculation we will assume idealized systems.  The B-coupler
scheme will divert 50\% of the incident light into the photometric
channels, and we will also assume the coupler is exactly balanced and 
loss-less.  Under these assumptions the output signals can be expressed as:

\begin{eqnarray}
P_1 & = & \frac{F_1}{2} \label{eq1a}\\
P_2 & = & \frac{F_2}{2} \label{eq1b}\\
I_1 & = & \frac{F_1}{4} + \frac{F_2}{4} + 2\sqrt{\frac{F_1}{4}\frac{F_2}{4}} \gamma(t) = \frac{F_1}{4} + \frac{F_2}{4} + \frac{1}{2}\sqrt{F_1 F_2} \gamma(t)
\label{eq1} \\
I_2 & = & \frac{F_1}{4} + \frac{F_2}{4} - 2\sqrt{\frac{F_1}{4}\frac{F_2}{4}} \gamma(t) = \frac{F_1}{4} + \frac{F_2}{4} - \frac{1}{2}\sqrt{F_1 F_2} \gamma(t)
\label{eq2}
\end{eqnarray} 

In the last two equations, 
$\gamma(t)$ is the mutual coherence function and encodes the
fringe visibility, the quantity we wish to measure.  Usually $\gamma$ is 
temporally modulated by adjusting the relative path lengths in the two
arms of the interferometer; this will be discussed further when we calculate
the signal-to-noise ratios for the different coupler designs.

The A-coupler will be defined so that the fraction $\alpha$ of $F_1$
is mixed with $\beta$ of $F_2$ for interferometric output channel
$I_1$, while the fraction $\beta$ of $F_1$ is mixed with fraction
$\alpha$ of $F_2$ for interference in $I_2$.  If the coupler is
loss-less, then $\beta=1-\alpha$, but I will keep the $\alpha$ and
$\beta$ notation for equation simplicity.  To summarize, 
the outputs of the A-coupler are given as:

\begin{eqnarray}
I_1 &= & \alpha F_1 + \beta F_2 + 2\sqrt{\alpha F_1 \beta F_2}\cdot \gamma(t) 
\label{eq3}\\
I_2 &= & \beta F_1 + \alpha F_2 - 2\sqrt{\alpha F_1 \beta F_2}\cdot \gamma(t) 
\label{eq4}
\end{eqnarray}

\subsection{Determining photometric signals with A-coupler}
\label{section:linear}

While determination of $F_1$ and $F_2$ is straightforward for the case of the 
B-coupler (see Eqs.\,\ref{eq1a} \& \ref{eq1b}), let us now discuss how to retrieve this same information using
the A-coupler.  We can split the $I_i$ signal into the sum of an incoherent and
a coherent term, $I=I_{\rm incoherent} + I_{\rm coherent}$.  The 
coherent part of this equation contains the interferometric ``fringe'' signal,
and is 
usually separated from the incoherent part by temporal modulation of 
$\gamma(t)$
at a frequency $\nu$ greater than $\frac{1}{\tau_0}$, where $\tau_0$ is the
coherence time of the atmosphere.  In order to estimate the photometric
flux from each telescope at the same time as we are detecting the fringe we
must be able to estimate the incoherent part of the $I$-channels.  
For the rest of this discussion, we will denote this incoherent component
of the $I$-channels as $\left< I_i \right>$, where the time-averaging
is sufficient to remove the fringe component but not atmospheric
fluctuations; this is equivalent to setting $\gamma(t)=0$ in Eqs. 
(\ref{eq3}) and (\ref{eq4}).  
The following SNR calculations for the A-coupler do assume one can
separate the coherent and incoherent parts perfectly.

In order to estimate $F_i$ using the A-coupler scheme we must use
linear combinations of $\left< I_1 \right>$ and $\left< I_2 \right>$.  
From inspection of Eqs. (\ref{eq3}) and (\ref{eq4}), we see that:
\begin{eqnarray}
F_1 & = &\frac{\alpha \ione - \beta \itwo}{\alpha^2 - \beta^2} 
\label{eq5}\\
F_2 & = &\frac{\beta \ione - \alpha \itwo}{\beta^2 - \alpha^2} 
\label{eq6}
\end{eqnarray}
Note that these equations become indeterminate when $\alpha=\beta$, 
corresponding to the balanced coupler case, $\alpha=\frac{1}{2}$.  
This tells us that
$\alpha$ must be significantly larger than $\beta$ in order to avoid
``amplifying'' the measurement uncertainty when estimating $F_i$.

\section{Signal-to-Noise Ratio Calculation}
In this section, I will compare the Signal-to-Noise Ratio (SNR) for
the proposed asymmetric coupler system (A-coupler) versus the traditional 
setup using a 50\% photometric tap and balanced couplers
for the interference (B-coupler).  

\subsection{Read-Noise Limit (Infrared)}
In the read-noise limited situation, we can imagine each output
channel having the same read-noise $\sigma_{\rm read}$.
Since the B-coupler scheme has twice
the number of output channels, we see that more read-noise is
introduced and we can expect degraded SNR.  
In the case of background-limited observations, the considerations
are similar to the read-noise dominated case, since
read-noise and background noise are both independent of signal
strength and average as the root of the number of independent
measurements.  However, the discussion is complicated since background
radiation can originate both before and after the combiner.  The 
background-limited case is discussed briefly in \S\ref{comparison}.

We proceed to calculate the SNR for the read-noise limited case.

\subsubsection{Photometric Channels}

The amplitudes of the photometric and interferometric signals were
determined in \S\ref{outputsigs} (Eq.\ref{eq1a}-\ref{eq6}).  For the
B-coupler, the noise in each photometric channel is obviously
$\sigma_{\rm read}$ per sample.  
For a total integration time of $N \, \Delta t$
(N samples each with integration time of $\Delta t$), it is trivial to
calculate the estimated signal-to-noise ratio (SNR) and this result
has been placed in Table\,\ref{table:snr_readnoise}.

For the A-coupler we form the linear combination described in
\S\ref{section:linear}, Eqs. (\ref{eq5}) and (\ref{eq6}).  In
estimating $F_i$ we must combine the errors from $\ione$ and $\itwo$
in quadrature, weighted by the appropriate factors.  Let us also
assume here we are averaging $N$ separate samples, each with
integration time of $\Delta t$, to take into account averaging over
the fringe modulation (see \S\ref{section:linear}).  In this case,
the SNR for $F_1$ can be written as:
\begin{eqnarray}
{\rm SNR}\, F_1 \equiv \frac{\rm Signal}{\rm Noise} & = &
  \frac{F_1 N \Delta t}{  \frac{ \sqrt{N}\sigma_{\rm read} \sqrt{\alpha^2 + \beta^2}  }{|\alpha^2 -\beta^2|} } \\
  &=& \frac{ |\alpha^2 -\beta^2|  F_1 \sqrt{N} \Delta t}
           {\sigma_{\rm read} \sqrt{\alpha^2 + \beta^2}} 
\label{eq7}
\end{eqnarray}
A similar derivation for $F_2$ completes
the first half of 
Table\,\ref{table:snr_readnoise}.

\subsubsection{Interferometric Channels}
It is more straightforward to calculate the SNR for the fringe amplitude.
The signal strength is simply equal to the coherent term of $I_i$, 
while the noise per sample is $\sigma_{\rm read}$.
 Eq. (\ref{eq3}) can be used to
calculate the signal-to-noise ratio for Fringe~1 as follows:
\begin{eqnarray}
{\rm SNR~Fringe~1} \equiv \frac{{\rm Coherent~part~of~}I_1}{
{\rm Read~Noise}} & = &
  \frac{2N \Delta t \sqrt{\alpha F_1 \beta F_2} |\gamma| }
{\sigma_{\rm read} \sqrt{ N }} \\
& = & \frac{2 \sqrt{N} \Delta t}{\sigma_{\rm read}}\sqrt{\alpha F_1 \beta F_2} 
\, |\gamma| 
\end{eqnarray}
Substituting $\alpha=\beta=\frac{1}{4}$, this result can be applied
for the case of B-coupler.  These results now appear in
Table\,\ref{table:snr_readnoise}.

\begin{table}[t]
\begin{center}
\caption{Signal-to-Noise Ratio Comparison: Read-Noise Limited
\label{table:snr_readnoise} }

\begin{tabular}{cccc}

\tableline
Quantity   &  B-Coupler  &  A-Coupler       & SNR Advantage\\
to be      &  SNR   &  SNR                  & of A-Coupler\\
Estimated  &               &                & over B-Coupler \\
\tableline
$F_1$ & $\frac{F_1 \sqrt{N} \Delta t}{2 \sigma_{\rm read}}$ &
 $  \frac{ |\alpha^2 -\beta^2|  F_1 \sqrt{N} \Delta t}
           {\sigma_{\rm read} \sqrt{\alpha^2 + \beta^2}}$  &
 $\frac{2  |\alpha^2 -\beta^2| }{ \sqrt{\alpha^2 + \beta^2}}$ \\ 
$F_2$ & $\frac{F_2 \sqrt{N} \Delta t}{2 \sigma_{\rm read}}$ &
 $  \frac{ |\beta^2 -\alpha^2|  F_2 \sqrt{N} \Delta t}
           {\sigma_{\rm read} \sqrt{\beta^2 + \alpha^2}}$  &
 $\frac{2  |\alpha^2 -\beta^2| }{ \sqrt{\alpha^2 + \beta^2}}$  \\
$I_1$ Fringe & $\frac{\sqrt{N} \Delta t}{2\sigma_{\rm read}}\sqrt{F_1 F_2}
 \, |\gamma| $&  
$\frac{2 \sqrt{N} \Delta t}{\sigma_{\rm read}}\sqrt{\alpha F_1 \beta F_2} 
\, |\gamma| $& $4\sqrt{\alpha \beta}$  \\

$I_2$ Fringe & $\frac{\sqrt{N} \Delta t}{2\sigma_{\rm read}}\sqrt{F_1 F_2} 
\, |\gamma| $& 
$\frac{2 \sqrt{N} \Delta t}{\sigma_{\rm read}}\sqrt{\alpha F_1 \beta F_2} 
\, |\gamma| $ & $4\sqrt{\alpha \beta}$\\
\end{tabular}
\end{center}
\end{table}

The SNR advantage of the A-coupler approach, 
$ \rm{SNRA} = \frac{\rm{A-Coupler~SNR}}{\rm{B-Coupler~SNR}}$,
 has been
included in the last column of Table\,\ref{table:snr_readnoise} and will
discussed in \S\ref{comparison}.

\subsection{Photon-Noise Limit (Visible)}
In the photon-limited regime, there is no explicit punishment for
using extra detectors or for multiple reads, 
and therefore we do not expect the A-coupler advantage 
to be as significant here.  
We will keep all the other notation the same.
The formulation for SNR for A- and B-couplers developed in the last section 
still applies, except we must use $\sigma_{\rm photon}$ 
instead of $\sigma_{\rm read}$.  For pure Poisson noise, the rms fluctuation
in the measurement of $F_1 N \Delta t$ photons is simply 
$\sqrt{ F_1 N \Delta t}$ photons.  We now proceed to create 
Table\,\ref{table:snr_photonnoise} containing the SNR for the 
B-coupler and A-coupler under photon-noise dominated conditions.

\subsubsection{Photometric Channels}
For the B-coupler, the noise in each photometric channel is
equal to the square root of the number of detected photons.  For a
total integration time of $N \, \Delta t$, the
estimated SNR is straightforward to 
to derive, and appears in Table\,\ref{table:snr_photonnoise}.

As before, use of the A-coupler requires the formation of linear
combinations (Eqs. \ref{eq5} and \ref{eq6}) of $\ione$ and $\itwo$ to
estimate $F_i$.  To correctly propagate the measurement uncertainties
of $\ione$ and $\itwo$ into our estimate of $F_i$, we combine the
Poisson errors from $\ione$ and $\itwo$ in quadrature, weighted by the
appropriate factors.  As before, we average $N$ separate samples, each
with integration time of $\Delta t$, to take into account averaging
over the fringe modulation.  In this case, the SNR
for $F_1$ can be written as:
\begin{eqnarray}
{\rm SNR}\, F_1 \equiv \frac{\rm Signal}{\rm Noise} & = &
  \frac{F_1 N \Delta t}
{  \frac{ {\left[  {\left(\alpha \sqrt{\ione N \Delta t}\right)}^2 +
                  {\left(\beta  \sqrt{\itwo N \Delta t}\right)}^2\right]}^{\frac{1}{2}} }
{|\alpha^2 -\beta^2|} } \\
 & = & \frac{F_1 \sqrt{N \Delta t}}
{  \frac{ \sqrt{  \alpha^2 (\alpha F_1 + \beta F_2) +
                  \beta^2  (\beta F_1 + \alpha F_2)}}
{|\alpha^2 -\beta^2|} } \\
  &=& \frac{ |\alpha^2 -\beta^2|  F_1 \sqrt{N \Delta t}}
{\sqrt{  \alpha^2 (\alpha F_1 + \beta F_2) +
                  \beta^2  (\beta F_1 + \alpha F_2)}}
\end{eqnarray}
I have not collected terms further in order to make it more apparent
that for the nominal case of a loss-less coupler ($\alpha+\beta=1$) and
for equal fluxes from each telescope ($\left< F_1 \right> = \left< F_2 \right>$), then $ {\rm SNR}\, F_1 = \frac{ |\alpha^2 -\beta^2|  \sqrt{F_1 N \Delta t}}
{\sqrt{  \alpha^2 +  \beta^2 }}$, which has the same
$(\alpha,\beta)$ dependences as for the read-noise limited case (see
Eq. \ref{eq7}).
The calculation for ${\rm SNR}\, F_2$ follows directly, and we place
these results in
Table\,\ref{table:snr_photonnoise}.

\subsubsection{Interferometric Channels}
As for the read-noise limited case, the SNR of the fringe amplitude
is more straightforward to derive than for the photometric signals.
The signal strength is simply equal to the {\em coherent term} of $I_i$, 
while the Poisson noise should be calculated based on
the {\em incoherent term} $\left<I_i\right>$, 
corresponding to the average flux.
Eq. (\ref{eq3}) can be used to
calculate the signal-to-noise ratio for Fringe~1 as follows:
\begin{eqnarray}
{\rm SNR~Fringe~1} \equiv \frac{{\rm Coherent~part~of~}I_1}{
{\rm Poisson~Noise~of~}\ione } & = &
  \frac{2N \Delta t \sqrt{\alpha F_1 \beta F_2} |\gamma| }
{\sqrt{ (\alpha F_1 + \beta F_2) N \Delta t}} \\
& = & \frac{2 \sqrt{N \Delta t} \sqrt{\alpha F_1 \beta F_2} |\gamma|}
{\sqrt{ (\alpha F_1 + \beta F_2)}} 
\label{eq8}
\end{eqnarray}
This result also applies for the B-combiner, substituting $\alpha=\beta=\frac{1}{4}$.  

As for the photometric SNR using the A-coupler, 
this result (Eq. \ref{eq8}) can be greatly 
simplified in the case of a loss-less 
combiner and equal telescope fluxes, i.e. 
${\rm SNR~Fringe~1} =2 \sqrt{F_1 N \Delta t} \sqrt{\alpha \beta} |\gamma| $,
which has the same $(\alpha,\beta)$ dependence as the read-noise
limited case.

\begin{table}[t]
\begin{center}
\caption{Signal-to-Noise Ratio Comparison: Photon-Noise Limited
\label{table:snr_photonnoise} }

\begin{tabular}{cccc}

\tableline
Quantity   &  B-Coupler  &  A-Coupler       & SNR Advantage\\
to be      &  SNR   &  SNR                  & of A-Coupler\\
Estimated  &               &                & over B-Coupler \\
\tableline
$F_1$ & $\sqrt{\frac{F_1 N \Delta t}{2}}$ &
$\frac{ |\alpha^2 -\beta^2|  F_1 \sqrt{N \Delta t}}
{\sqrt{  \alpha^2 (\alpha F_1 + \beta F_2) +
                  \beta^2  (\beta F_1 + \alpha F_2)}} $ &
 $\frac{\sqrt{2}  |\alpha^2 -\beta^2| }{
\sqrt{ \alpha^2\left(\alpha + \beta \frac{F_2}{F_1}\right) + 
       \beta^2 \left(\beta + \alpha\frac{F_2}{F_1}\right)}}$\\
$F_2$ & $\sqrt{\frac{F_2 N \Delta t}{2}}$ &
$\frac{ |\beta^2 -\alpha^2|  F_2 \sqrt{N \Delta t}}
{\sqrt{  \beta^2 (\alpha F_1 + \beta F_2) +
                  \alpha^2  (\beta F_1 + \alpha F_2)}} $ &

 $\frac{\sqrt{2}  |\alpha^2 -\beta^2| }{
\sqrt{ \alpha^2\left(\alpha + \beta \frac{F_1}{F_2}\right) +
       \beta^2 \left(\beta + \alpha\frac{F_1}{F_2}\right)}}$  \\
$I_1$ Fringe & 
$\frac{\sqrt{ N \Delta t~ F_1 F_2 }  |\gamma|}{\sqrt{F_1 + F_2}}$ &
$ \frac{2 \sqrt{N \Delta t~ \alpha \beta F_1 F_2} |\gamma|}
{\sqrt{ (\alpha F_1 + \beta F_2)}}$  & 
$2\sqrt{\alpha\beta}\sqrt{\frac{1 + \frac{F_2}{F_1}}{\alpha + \beta \frac{F_2}{F_1}}}$
\\
$I_2$ Fringe & 
$\frac{\sqrt{ N \Delta t~ F_1 F_2 }  |\gamma|}{\sqrt{F_1 + F_2}}$ &
$ \frac{2 \sqrt{N \Delta t~ \alpha \beta F_1 F_2} |\gamma|}
{\sqrt{ (\beta F_1 + \alpha F_2)}}$  &
$2\sqrt{\alpha\beta}\sqrt{\frac{1+ \frac{F_1}{F_2}}
{\alpha + \beta \frac{F_1}{F_2}}}$ \\
 & 
\end{tabular}
\end{center}
\end{table}

\subsection{Comparison}
\label{comparison}
In order to facilitate comparison between the performance of the A-coupler 
and the B-coupler, we will make a few simplifying assumptions.  
We will assume that $\left<F_1\right> = \left<F_2\right>$  
and that the A-coupler is loss-less ($\alpha+\beta=1$).
In addition, we will introduce a new quantity, $A=\frac{\alpha}{\beta}$, the 
``Asymmetry Factor,'' defined greater than or equal to unity.  
Note that for a balanced system, the Asymmetry Factor is equal to Unity.

For this simplified configuration, the SNRs for the two photometric
determinations are identical, as are the SNRs for Fringes~1 and
Fringes~2.  Therefore, we will stop differentiating between the two
and speak only of the SNR of the photometric signal (i.e., the flux
amplitude) and of the interferometric signal (i.e., the fringe
amplitude).  Table\,\ref{table:final} contains the SNR advantage of
the A-coupler over the B-coupler as a function for Asymmetry Factor
($A$) for the read-noise and photon-noise limit cases based on results
contained in the Tables\,\ref{table:snr_readnoise} and
\ref{table:snr_photonnoise}.  Note that the SNR advantage between the
read-noise-limited and photon-limited regimes differs by exactly
$\sqrt{2}$ for both the flux and the fringe amplitude estimators.

It is interesting to note that the background-limited case falls
somewhere in-between these two limits.  The photometric channels and
the non-unity coupling efficiency will introduce background radiation
at the temperature of the fiber optic components.  For a free-space
system, this is equivalent to the fact that you can not ``split-off''
a fraction of a beam without introducing new lines-of-sight to the
detector not coupled to the sky; thus additional background degrades
the measurement.  However if the combiner is cryogenically cooled, the
background-limited case will be identical to the photon-noise limited
case.  Alternatively, if emission from the combiner dominates
sky+telescope emission then the background-limited case will share the
same SNR as the read-noise limited case.  In practice, the result will
intermediate.

If we assume the SNRs of the flux amplitude and fringe amplitude are
equally important for estimating the fringe visibility (or contrast),
which is probably only true for relatively bright sources, 
then we can estimate the combined SNR Advantage (SNRA) of
the A-coupler over the B-coupler by applying equal weights and adding the
errors in quadrature. Mathematically, this is done as follows:
\begin{equation}
SNRA_{\rm Visibility} \equiv \frac{\sqrt{2}}{\sqrt{ 
\frac{1}{ {SNRA_{\rm Flux}}^2} + \frac{1}{ {SNRA_{\rm Fringe}}^2}}}
\end{equation}

\begin{figure}[thb]
\begin{center}
\includegraphics[angle=90, width=4in]{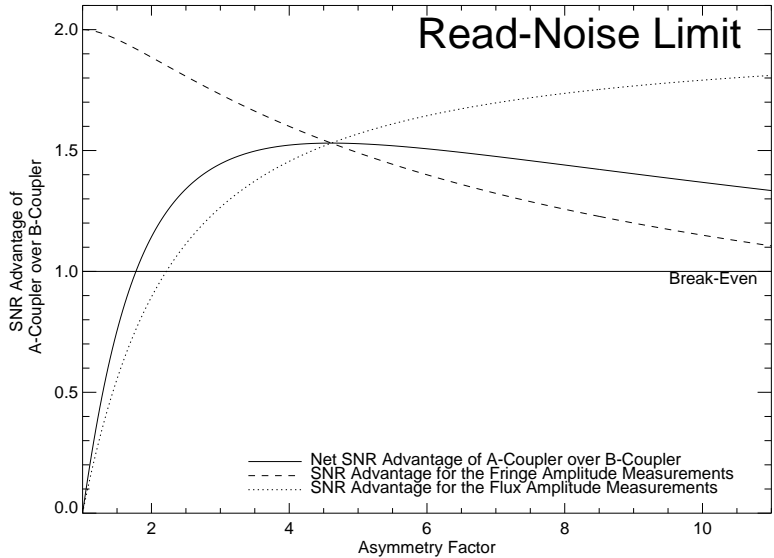}
\caption{Signal-to-Noise Ratio Advantage for Asymmetric-Coupler over
Balanced Coupler, as a function of the A-coupler Asymmetry Factor, assuming
read-noise limited performance.
\label{fig:snra_readnoise}}
\end{center}
\end{figure}

\begin{figure}[thb]
\begin{center}
\includegraphics[angle=90, width=4in]{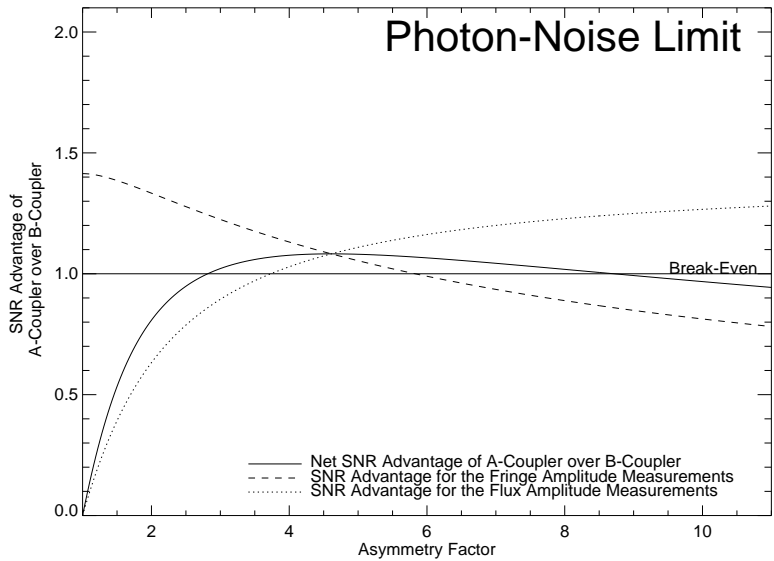}
\caption{Signal-to-Noise Ratio Advantage for Asymmetric-Coupler over
Balanced Coupler, as a function of the A-coupler Asymmetry Factor, assuming
photon-noise limited performance.
\label{fig:snra_photonnoise}}
\end{center}
\end{figure}

These results are shown graphically in 
Figures\,\ref{fig:snra_readnoise} and \ref{fig:snra_photonnoise}.
In the read-noise limited case, we see that the A-coupler design is
superior for Asymmetry factors greater than 1.78 and less than 25.63, with
maximum improvement of 53\% for $A=4.61$.  In the photon-noise
limit, the A-coupler is only slightly superior between Asymmetry
factors of 2.82 and 8.71, with a maximum improvement of 8.2\%, 
also at $A=4.61$.  

Strictly speaking, the photometric signal can be averaged for longer
than the fringes since the fringes must be modulated faster than the
atmospheric time scale.  This means that the SNR of the B-coupler
setup can be improved by diverting a smaller percentage of the flux to
the photometric taps; the optimal ratio would depend on various
instrumental and observing factors, such as the inherent fringe
visibility of your source and the temporal coherence time of the
atmosphere.  However by the same argument, the SNR of the A-coupler
setup can be also improved by using a smaller asymmetry factor.  These
details have been neglected in the above calculation, but are not
expected to qualitatively change the results.

\begin{table}[t]
\begin{center}
\caption{Summary: Signal-to-Noise Ratio as Function of Asymmetry Factor~A
\label{table:final} }

\begin{tabular}{ccc}

\tableline
Quantity   & \multicolumn{2}{c}{SNR Advantage of} \\
to be      & \multicolumn{2}{c}{A-Coupler over B-Coupler} \\
Estimated  & [Read-Noise Limit] & [Photon-Noise Limit] \\
\tableline
Flux Amplitude & $2 \frac{(A - 1)}{\sqrt{A^2+1}}$    &
                 $\sqrt{2}\frac{(A - 1)}{\sqrt{A^2+1}} $\\
Fringe Amplitude & $4\frac{\sqrt{A}}{A+1}$& 
 $2\sqrt{2}\frac{\sqrt{A}}{A+1}$\\
\end{tabular}
\end{center}
\end{table}
 

\section{Discussion}
We see that the A-coupler offers a significant improvement for the
read-noise limited case, and a slight improvement in the photon-noise
limit.  This method will work with a FLUOR-type system where beam
combination occurs in a fiber optic coupler or integrated optic
\citep{berger99}.  
Fringe-tracking systems can benefit from this
scheme as well, since the mean signal in each interferometric channel
can be separately determined by binning all data in each
fringe-modulation period, but this does require {\em both}
interferometric channels to be measured simultaneously. 
Asymmetric beam combination is also useful when there is no spatial filtering.
Other effects can cause modulation of the beam intensities from telescopes,
including scintillation and guiding errors, and the methods outlined here
can be used to correct for these effects.

For
interferometric arrays with three or more telescopes, other
possibilities exist for determining the individual telescope
photometric strengths.  The A-coupler scheme can be generalized for
``all-in-one'' combiners \citep[e.g.,][]{coast94a}, for instance using
four asymmetric couplers allows one to differentiate the four input
signals using the four output signals.  Interestingly for the
three-telescope (or more) pair-wise combination scheme, the photometric signals
can be determined even with standard, symmetric couplers using linear
combinations similar to the ones derived here.

Another advantage to the A-coupler design is decreased complexity of
the combiner, although the data analysis will be more
complicated.  This is especially important in the visible, because of the
expensive cost of high quantum efficiency, photon-counting detectors,
such as avalanche photo-diodes (APDs).  One apparent disadvantage to an
A-coupler design is that one can no longer subtract the two
interferometric signals to remove ``common-mode'' fluctuations such as
scintillation.  However, when the fringes are modulated faster than
scintillation time scales, these variations can be completely corrected
by forming the appropriate linear combinations as already discussed herein.

One caveat should be mentioned here.  When spatial filtering occurs
after (free-space) 
beam combination \citep[e.g.,][]{colavita99}, these methods may
not work well.  The reason is that optical misalignments leading to
the single-mode fibers can cause decorrelations between the coupling
efficiencies on either side of the free-space beamsplitter.  For
instance, a wavefront tilt introduced by the atmosphere could cause an
increase in coupling efficiency for one fiber, but a decrease for the
other fiber if the optical axes of both fibers are not aligned
identically with the telescope axis.  Hence, the photometric
estimators will not be a fixed linear combination of the
interferometric outputs, degrading the utility of the technique.

An experiment is underway at the Harvard-Smithsonian Center for
Astrophysics to test these ideas, using a 780\,nm asymmetric coupler,
asymmetry factor 2.3.  If lab tests prove satisfactory, the
equipment will be incorporated into IOTA \citep{carleton94}, utilizing much
of the existing visible light detection system for
measurements of stellar diameters.
The ultimate goal of this experiment is to yield high
precision diameters measurements ($<<$1\%), allowing the pulsation
amplitude to be measured for calibration of the Cepheid distance
scale.

\section{Conclusions}
Asymmetric beam combination in optical interferometry 
can improve sensitivity while reducing combiner complexity and cost, 
especially in the read-noise limited regime encountered
for most infrared observing.    

\acknowledgments
I appreciate discussions with Rafael Millan-Gabet, Jean-Phillipe
Berger, Wes Traub, and other IOTA group members, and thank Peter
Tuthill and David Mozurkewich for comments on the manuscript.  Nat
Carleton pointed out how photometric correction for a three-telescope,
pair-wise combination scheme can be achieved with symmetric couplers,
using appropriate linear combinations of the output channels.  
After the manuscript was submitted, the author was informed that a similar
idea was proposed in 1995 by Jean Gay of Observatoire de la Cote d'Azur in 
Nice.
JDM
acknowledges support from a Center for Astrophysics Fellowship at the
Harvard-Smithsonian Center for Astrophysics.

\bibliographystyle{apj}
\bibliography{apj-jour,Coupler,Thesis}



\clearpage








\clearpage

\end{document}